\documentstyle[12pt]{article}

\begin{document}
\title{\textbf{Spectrum of kinematic fast dynamo operator in Ricci flows}} \maketitle
{\sl \textbf{L.C. Garcia de Andrade}\newline
Departamento de F\'{\i}sica
Te\'orica-IF\newline
Universidade do Estado do Rio de Janeiro\\[-2mm]
Rua S\~ao Francisco Xavier, 524\\[-2mm]
Cep 20550-003, Maracan\~a, Rio de Janeiro, RJ, Brasil\\[-3mm]
\vspace{0.01cm} \newline{\bf Abstract} \paragraph*{}Spectrum of kinematic fast dynamo operators in Ricci compressible flows in Einstein 2-manifolds is investigated. A similar expression, to the one obtained by Chicone, Latushkin and Montgomery-Smith (Comm Math Phys (1995)) is given, for the fast dynamo operator. The operator eigenvalue is obtained in a highly conducting media, in terms of linear and nonlinear orders of Ricci scalar and diffusion-free limit. Spatial 3-Einstein manifold section of Friedmann-Robertson-Walker (FRW) cosmology is obtained in the limit of ideal plasma. Since only two dimensional Riemannian manifolds of negative curvature may support fast dynamo action, here only inflationary phase of the universe, of negative cosmological constant $({\Lambda})$ in Ricci-Einstein flows may support dynamo action. When $({\Lambda}\ge{0})$ magnetic field decays. As in Latushkin and Vishik (Comm Math Phys (2003)) Lyapunov exponents in kinematic dynamos are investigated. Since positive curvature scalars are preserved under Ricci flow, fast dynamos are also preserved. Bounds on cosmological model due to Vishik anti-fast dynamo theorem are also discussed.}
\newpage
\section{Introduction}
Investigations in the Riemannian geometry of magnetic dynamos ranged from the early investigations of Arnold, Zeldovich, Ruzmaikin and Sokoloff \cite{1} to the more recently papers by Chiconne, Latushkin and their group \cite{2} on the fast dynamo existence. Investigation of Riemannian geometry applications to plasma dynamos and anti-fast dynamo Vishik's theorem \cite{3} has been performed by the Garcia de Andrade \cite{4}. Chicone et al \cite{2} have also shown that the fast dynamo operator spectrum for an ideally conducting fluid and the spectrum of the group acting on the associated compact Riemannian manifold. Yet more recently Latushkin and Vishik \cite{5} investigated Lyapunov exponents in kinematic dynamos.  On the more topological and dynamical system settings, Young and Klapper \cite{6} has investigated the topological entropy systems and dynamo action. In this last paper Young and Klapper used the same concept of topological entropy used recently, by Fields medalist Grisha Perelman \cite{7} to prove the long standing Poincare conjecture. In his important proof, Perelman has made used the concept of Ricci flows, proposed by Hamilton in 1982 \cite{8}. In this paper a proof is given of the following theorem:\newline
\textbf{Theorem}: Let $\cal{M}$ be a two dimensional Riemannian manifold $(\cal{M},\textbf{g})$, not necessarily closed, endowed with a metric $\textbf{g}(t)$, is given in the interval $t\in{[a,b]}$ in the field of real numbers $\textbf{R}$ of an Einstein 3-manifold and a Ricci flow given by the equation\newline
\textbf{Definition}:
\begin{equation}
\frac{{\partial}\textbf{g}}{{\partial}t}=-2\textbf{Ric}
\label{1}
\end{equation}
where $\textbf{Ric}$ represents the Ricci tensor with components $R_{ij}$ on a chart $\cal{U}$, submanifold of the tangent space $T\cal{M}$ with coordinates $(i,j=1,2)$, thus the kinematic fast dynamo operator (where back reactions due to Lorentz forces are negligeble), in ideally highly conducting fluid is
\begin{equation}
{\Gamma}_{\eta}=\frac{{\partial}}{{\partial}t}=(-div{\textbf{v}}+\textbf{Ric}+{\eta}{\Delta})\label{2}
\end{equation}
while the its spectrum eigenvalues are given by
\begin{equation}
{\lambda}_{\eta}= [-(R+\frac{1}{2}{\theta}-{\eta})\pm{\sqrt{-\frac{3}{4}{\theta}^{2}+{\eta}{\theta}(1-\frac{R}{\theta})-\frac{3}{4}{\theta}^{2}(1+\frac{R}{\theta})-{\eta}^{2}}}]
\label{3}
\end{equation}
where R is the trace of the $\textbf{Ric}$ and ${\lambda}$ is the growth rate usually found in dynamo theory, and the magnetic field obeys the rule $|\textbf{B}|\approx{e^{{\lambda}t}}$. In general where the diffusion constant $\eta$ does not vanish the dynamo operator is written as ${\Gamma}_{\eta}\textbf{B}=curl(\textbf{v}{\times}\textbf{B})+{\eta}{\Delta}\textbf{B}$.
The absence of shear and vorticity traces are due to the fact that they vanish in geodesic flows. Here, $\textbf{g}$ is the Riemann metric, over manifold $\cal{M}$, and the parameter t in the Riemann metric $\textbf{g}(t)$, is given in the interval $t\in{[a,b]}$ in the field of real numbers $\textbf{R}$. Several interesting corollaries can be obtained from the eigenvalue expression above. The first is the case of diffusion free flow, where the diffusion parameter vanishes. In this case the eigenvalue turns out to be
\begin{equation}
{\lambda}_{0}= [-(R+\frac{1}{2}{\theta})\pm{\sqrt{-\frac{3}{4}{\theta}^{2}-\frac{3}{4}{\theta}^{2}
(1+\frac{R}{\theta})}}]
\label{4}
\end{equation}
Since in cosmological models post-inflationary era, the Ricci curvature is given by
\begin{equation}
R={\rho}+{\theta}\label{5}
\end{equation}
where ${\rho}$ is the universe matter density. Here the universe is considered pressure-free, and only cosmological constant plays its role. Substitution of the expression (\ref{5}) into (\ref{6}) reduces it to
\begin{equation}
{\lambda}_{\eta}= \frac{1}{2}[(-3R+{\rho}+2{\eta})\pm{i(\sqrt{{7}{\rho}^{2}
+{4}{\rho}^{2}(1-2\frac{R}{\rho})+4{\eta}^{2}})}]
\label{6}
\end{equation}
Since the fast dynamo criteria \cite{9} states that the existence of fast dynamo is given by
\begin{equation}
lim_{{\eta}\rightarrow{0}}\textbf{Re}{\lambda}_{\eta}>{0} \label{7}
\end{equation}
application of this criteria to the real part of the complex eigenvalue formula (\cite{6})
\begin{equation}
\textbf{Re}{\lambda}_{\eta}= \frac{1}{2}(-3R+{\rho})
\label{8}
\end{equation}
shows that, since the cosmic density ${\rho}$ is positive, for the Ricci cosmological flow to support a fast dynamo action, either the Ricci curvature scalar $Tr\textbf{Ric}<0$, or that the inequality, $-3{\theta}\ge{2{\rho}}$. The first case coincides with the Chicone and Latushkin result that scalar curvature of the dynamo curved 2-surface be negative as the case of fast dynamo action in geodesic flows of negative constant Riemannian curvature, the so-called Anosov flows \cite{10}. Since the matter density is positive, the last relation indicates that the expansion has to be negative, or that the universe undergoes a contracting phase to support dynamos in Ricci flows. This constraint is actually very severe, since the stretching of magnetic field by the plasma flow in de Sitter model, allows for the presence of dynamo action as well. Note that, otherwise, the $\textbf{Re}{\lambda}_{\eta}$ magnetic growth rate is negative, thus
\begin{equation}
\textbf{Re}{\lambda}_{\eta}= \frac{1}{2}(-3R+{\rho})\le{0}
\label{9}
\end{equation}
implies that
\begin{equation}
{\rho}\le{3R}
\label{10}
\end{equation}
This shows that the Ricci curvature scalar represents an upper bound to the expansion of the universe two dimensional spatial section. This slows down the universe expansion and the dynamo action cannot be supported in this case. Expansion, of course still occurs since from expression (\ref{10}) one notes that the expansion ${\theta}$ and curvature R, can be both positive. There is also the possibility of marginal dynamos $({\lambda}=0)$, but from expression (\ref{5}) this is the Einstein static universe \cite{11}, where expansion ${\theta}$ vanishes, since $R={\rho}$. After this physical interpretation of the cosmic fast dynamo operator, in the next section one shall prove the basic theorem $I.1$. Since this is a long demonstration, we left it to a whole section.
\section{Cosmic dynamo Ricci flow theorem: A simple proof}
In the beginning of the proof, we show that the de Sitter spatial section cosmology, actually is a natural chaotic flow with Lyapunov exponent leading to chaotic dynamo flows, in the absence of diffusion. On a local chart, of the Riemannian manifold tangent space, expression (\ref{1}) can be expressed as \cite{7}
\begin{equation}
\frac{{\partial}{g_{ij}}}{{\partial}t}=-2{R_{ij}}
\label{11}
\end{equation}
where $\textbf{Ric}$, is the Ricci tensor, whose components $R_{ij}$. From this expression, one defines the eigenvalue problem as
\begin{equation}
R_{ij}{\chi}^{j}={\lambda}{\chi}_{i}\label{12}
\end{equation}
where $(i,j=1,2)$. Substitution of the Ricci flow equation (\ref{2}) into this eigenvalue expression and cancelling the eigendirection ${\chi}^{i}$ on both sides of the equation yields
\begin{equation}
\frac{{\partial}g_{ij}}{{\partial}t}=-2\lambda{g_{ij}}
\label{13}
\end{equation}
Solution of this equation yields the de Sitter-Lyapunov expression for the metric
\begin{equation}
g_{ij}=exp{[-2\lambda{t}]}{\delta}_{ij}
\label{14}
\end{equation}
where ${\delta}_{ij}$ is the Kroenecker delta. Note that in principle if ${\lambda}\le{0}$ the metric grows without bounds, and in case it is negative it is bounded as $t\rightarrow{\infty}$. Recently, Thiffeault \cite{12} has used a similar Lyapunov exponents expression in Riemannian manifolds to investigate chaotic flows, without attention to dynamos or Ricci flow. These exponents are obtained in the metric as
\begin{equation}
\textbf{g}={\Lambda}_{11}\textbf{e}_{1}\textbf{e}_{1}+{\Lambda}_{22}\textbf{e}_{2}\textbf{e}_{2}
\label{15}
\end{equation}
Thus one has proven the following lemma:\newline
\textbf{Lemma II.1}:\newline
If ${\lambda}_{i}$ is an eigenvalue spectra of the $\textbf{Ric}$ tensor, the finite-time Lyapunov exponents spectra is given by
\begin{equation}
{\lambda}_{i}=-{\gamma}_{i}\le{0}\label{16}
\end{equation}
In the next section I shall use this argument to work with the de Sitter metric
\begin{equation}
ds^{2}=-dt^{2}+e^{{\Lambda}t}(dx^{2}+dy^{2})\label{17}
\end{equation}
In the next and last section, one shall be concerned with the proof of the above theorem through the self-induction equation in Ricci flows in Einstein 2-manifolds. Here, one shall use the formalism of dynamo theory as in the book by Arnold and Khesin \cite{10} in chaotic flows and Anosov geodesic flows \cite{9}. Actually in the Einstein manifold case below, the space of constant negative curvature induces negative cosmological constant. This is in agreement with the fact that only spaces of constant negative curvature in two dimensions, support dynamo action \cite{15}. Let us consider first the kinematic self-induced equation in Euclidean three-dimensional space $\textbf{E}^{3}$, as dynamo equation. In the mathematician notation, Arnold writes down the Poisson bracket
\begin{equation}
\{v,B\}=-curl(v{\times}B)\label{18}
\end{equation}
between the flow v and the magnetic field B. The self-induced magnetic field equation reads
\begin{equation}
\frac{{\partial}B}{{\partial}t}=-\{v,B\}+div(v)B+{\eta}{\Delta}B\label{19}
\end{equation}
where ${\eta}$ is the plasma resistivity and ${\Delta}:={\nabla}^{2}$ is the Laplacian operator. In Chicone et al work the Riemannian manifold was confined to incompressible dynamo flows, where the second term on the RHS of expression (\ref{19}) vanishes, and along with the divergence-free condition of the magnetic field
\begin{equation}
div B=0\label{20}
\end{equation}
they form a solenoidal vector field in Riemannian manifold. Here incompressibility of the flow is not assumed and only solenoidal property of the magnetic field is kept. As one shall see, is exactly this compressible flow term that is responsible for the introduction of the Ehlers-Sachs optical parameters of shear, vorticity and expansion of the lower dimensional universe. Earlier a two dimensional model of a dynamo flow has been numerically tested, by Otani \cite{13}. More recently, an example of a two dimensional Moebius dynamo flow, has also been obtained by Shukurov, Stepanov and Sokoloff \cite{14} to model the Perm dynamo experiment. Let us now consider, the diffusive term given by
\begin{equation}
{\Delta}\textbf{B}= \frac{1}{\sqrt{g}}{\partial}_{i}[\sqrt{g}g^{ij}{\partial}_{j}\textbf{B}]\label{21}
\end{equation}
which expanded in the frame ${\textbf{e}}_{i}$ with
\begin{equation}
\textbf{B}=B^{i}\textbf{e}_{i}\label{22}
\end{equation}
yields
\begin{equation}
{\Delta}\textbf{B}= [{g}^{ij}{\partial}_{i}{\partial}_{j}B^{p}+B^{k}[{\partial}_{i}{{\gamma}^{p}}_{jk}g^{ij}+
{{\gamma}^{l}}_{jk}{{\gamma}^{p}}_{il}g^{ij}]+[{{\gamma}^{p}}_{jk}g^{ij}{\partial}_{i}B^{k}]]
\textbf{e}_{p}\label{23}
\end{equation}
Here ${{\gamma}^{l}}_{jk}$ represents the Ricci rotation coefficients (RRCs) analogous to the Riemann-Christoffel symbols, which is defined by
\begin{equation}
{{\partial}_{k}}\textbf{e}_{i}={{\gamma}_{ki}}^{j}\textbf{e}_{j}\label{24}
\end{equation}
The Christoffel symbols
\begin{equation}
{{\Gamma}^{i}}_{jk}=g^{il}[{{\partial}_{j}}g_{kl}+{{\partial}_{k}}g_{jl}-
{\partial}_{l}g_{jk}]\label{25}
\end{equation}
do not appear in the computations, since we have assumed, that the trace of the Christoffel symbols vanish. To complete the derivation of the self-induction equation it remains to obtain the diffusion free part of the self-induced equation above , which in general curvilinear coordinates ${x^{i}}\in{\cal{U}}_{i}$, of the sub-chart $U_{i}$ of the manifold, in the rotating frame reference of the flow $\textbf{e}_{i}$, reads
\begin{equation}
\frac{d\textbf{B}}{dt}=<{\textbf{B}},{\nabla}>\textbf{v}-(div\textbf{v})\textbf{B}\label{26}
\end{equation}
Before this derivation, let us now introduce the Ricci tensor into play, by considering the following trick
\begin{equation}
\frac{d}{dt}[g^{il}g_{lk}]=\frac{d}{dt}[{{\delta}^{i}}_{k}]=0\label{27}
\end{equation}
which can be applied to the expression
\begin{equation}
\frac{d\textbf{B}}{dt}= \frac{d}{dt}(g^{ik}B_{k}\textbf{e}_{i})\label{28}
\end{equation}
to obtain
\begin{equation}
\frac{d}{dt}{(g^{ik}B_{k})}= \frac{d}{dt}(g^{ik})B_{k}+g^{ik}\frac{d}{dt}{B_{k}}\label{29}
\end{equation}
Now by making use of the Ricci flow equation above into this last expression, yields
\begin{equation}
\frac{d}{dt}{g^{ik}B_{k}}= -2R^{ik}B_{k}+g^{ik}\frac{d}{dt}{B_{k}}-div\textbf{v}B^{i}\label{30}
\end{equation}
Now let us consider that the Ehlers-Sachs optical scalar of the fluid appears in the gradient of the flow expression such as
\begin{equation}
({\nabla}\textbf{v})_{pl}={\nabla}_{p}v_{l}={\Omega}_{pl}+{\sigma}_{pl}-
\frac{1}{3}{\theta}g_{lp}-A_{p}v_{l}
\label{31}
\end{equation}
where A is the acceleration of the flow, and ${\nabla}_{p}$ is the covariant derivative operator of Riemannian geometry. Taking the trace of this expression, one obtains the $div\textbf{v}$ expression as given by the gradient strain
\begin{equation}
Tr({\nabla}\textbf{v})=div\textbf{v}={\partial}_{p}v^{p}={\sigma}-
{\theta}
\label{32}
\end{equation}
where, ${\sigma}:=Tr({\sigma}_{ij})$ is the trace of the shear tensor ${\sigma}_{ij}$ while, ${\theta}$ is the expansion of the flow. In cosmology they vanish and the acceleration is orthogonal to velocity, that is the reason they shall dissappear throughout from the rest of the paper computations. In the language of dynamo theory, this scalar represents the stretching of the dynamo flow. The magnetic field here is stretched by a cosmological Ricci two-dimensional flow, what Arnold has called a particle stretching in the flow. From evolution of the reference frame
\begin{equation}
\frac{d\textbf{e}_{i}}{dt}={{\omega}_{i}}^{j}\textbf{e}_{j}\label{33}
\end{equation}
where we have used the following expression
\begin{equation}
{{\partial}_{k}}\textbf{e}_{i}={{\Gamma}_{ki}}^{j}\textbf{e}_{j}\label{34}
\end{equation}
Introduction of the diffusion through the term ${\Delta}\textbf{B}=-curl(curl\textbf{B})$ and the assumption that the $curl\textbf{B}=\textbf{B}$ yields The eigenvalue problem
\begin{equation}
{\Gamma}_{\eta}\textbf{B}={\lambda}\textbf{B}\label{35}
\end{equation}
yields the following algebraic second order eigenvalue equation
\begin{equation}
{\lambda}^{2}+2[R+\frac{{\theta}}{2}-{\eta}]{\lambda}+[R^{2}+{\theta}^{2}+2({\eta}-{\theta})R]=0
\label{36}
\end{equation}
where one has consider that in Einstein constant curvature manifold the components of the Ricci tensor components coincide as $R_{11}=R_{22}=R$. Note that a simple solution yields the result (\ref{2}) for the eigenvalue ${\lambda}_{0}$ above and the theorem is proved. Just for comparison with the Chiconne and Latushkin \cite{15} case and convenience of the reader, one repeats here their eigenvalue expression
\begin{equation}
{\lambda}_{\eta}= \frac{1}{2}[-{\eta}(1+{\kappa}^{2})+\sqrt{-4{\kappa}+{\eta}(1-{\kappa}^{2})}]
\label{37}
\end{equation}
Where ${\kappa}$ is the surface curvature scalar similar to the Ricci scalar, $Tr\textbf{Ric}$ used in this paper. This is in agreement with Woolgar \cite{15} argument that in Ricci flows, Einstein gravity is similar, but cosmology departures from the more traditional relativistic cosmology. One notes that in the absence of diffusion $\eta$ vanishes, and this expression reduces to
\begin{equation}
{\lambda}_{0}= \frac{1}{2}[\sqrt{-4{\kappa}}]
\label{38}
\end{equation}
This result is similar to the one obtained in the last section, since it leads to the sam fact that fast dynamos are naturally supported in compact two-dimensional Riemannian  manifolds of constant curvature. A corollary can be obtained by investigating the eigenvalue degeneracy and the discriminant ${\Delta}$ of the second order equation given by
\begin{equation}
{\Delta}=11{\rho}-8R
\label{39}
\end{equation}
 Thus, the degenerate eigenvalues implies that ${\rho}=0.72R=0.72{\Lambda}$. This shows that matter density are weaker than the cosmological constant and Ricci curvature, which would be an unphysical situation that the universe would be closed again, very far from the actual flat universe. This contradicts present cosmological Hubble satellite observation, which shows that based in inflationary models the actual universe would be flat, thus the degenerate eigenvalues cannot appear in the fast cosmic dynamo in Ricci flows.\newline
 \textbf{Corollary II.1}:\newline
Expansion of the non-relativistic Ricci flow two dimensional universe cannot support dynamo action, while a contracting phase of de Sitter-like model can. Since, as shown by R S Hamilton \cite{8}, the curvature scalar is preserved under Ricci flow, here one has proven the following lemma:\newline
\textbf{Lemma II.2}:\newline
Since the positive curvature scalar is preserved under the Ricci flow, and the fast dynamo action is supported in this case, the fast dynamo action is preserved under the Ricci flow.\newline
When the Ricci flows obey the Einstein 2-manifold  condition
\begin{equation}
\textbf{Ric}={\Lambda}\textbf{g}
\label{40}
\end{equation}
The value of the Ricci curvature R should be simply substituted by the cosmological constant. \newline
Taking into account the magnetic energy ${\epsilon}$ as
\begin{equation}
{\epsilon}=\int{B^{2}{\mu}}
\label{41}
\end{equation}
one is able to simply shown that in terms of the 2D Riemann metric components
\begin{equation}
{\epsilon}=\int{B^{i}g_{ij}B^{j}{\mu}}
\label{42}
\end{equation}
Here ${\mu}$ represents the volume form. Since, by definition fast dynamo action corresponds to the growth of magnetic energy in time as $\frac{{\partial}{\epsilon}}{{\partial}t}\ge{0}$, this amount has to be computed by performing the partial time derivative of the expression (\ref{38}). Actually the equal sign in the last condition represents the lower limit of marginal dynamos, where the magnetic energy integral remains constant. This computation yields
\begin{equation}
\frac{{\partial}{\epsilon}}{{\partial}t}=\frac{{\partial}[\int{B^{i}g_{ij}B^{j}{\mu}}]}{{\partial}t}
\label{43}
\end{equation}
Expansion of the RHS of this expression shows clearly now where the Ricci flow eigenvalue effect is going to appear. A simple computation, shows that the energy integral confirms the dynamo action. Throughout the paper the diffusion term was not explicitly computed since because we use the limit of diffusion free to check for the presence of slow dynamos, which seems not exist globally in the universe. Note that the in the de Sitter case the magnetic can be written as
\begin{equation}
B^{i}={B^{i}}_{0}e^{[2{Tr\textbf{Ric}}-div\textbf{v}]t}
\label{44}
\end{equation}
which is equivalent to the most familiar cosmologists expression
\begin{equation}
B^{i}={B^{i}}_{0}e^{[2{\Lambda}-{\theta}]t}
\label{45}
\end{equation}
Substitution of this expression into the magnetic energy one, is possible to obtain the result
\begin{equation}
{\epsilon}\approx{e^{2[2{\Lambda}-{\theta}]t}}
\label{46}
\end{equation}
which confirms the growth rate of magnetic energy of the fast dynamo as long as the upper bound for the universe expansion be fullfilles as ${\Lambda}\ge{\frac{1}{2}{\theta}}$. Anti-de Sitter effective \cite{16} spacetime, of course contributes to slow down magnetic field as negative exponents contributes to the decay of magnetic field in the effective universe. From the mathematical point of view, one may write expression (\ref{46}) in the operator format as
\begin{equation}
B^{i}={B^{i}}_{0}e^{t{\Gamma}_{\eta}}
\label{47}
\end{equation}
where the term $(e^{t{\Gamma}_{\eta}})_{t\in{\textbf{R}}}$ represents the group of Lie derivatives of the kinematic fast dynamo operator ${\Gamma}_{\eta}$ in the limit of diffusion free media and compressible flows. Another basic distinction between the Chicone-Latushkin work and the present one is that, here the flows are compressible instead the divergenge-free kinematic fast dynamo flows in compact Riemannian manifolds.
When one applies the anti-fast dynamo theorem \cite{3}, as the negative limit of the Lyapunov exponent of the kinematic dynamo operator ${\Gamma}_{\eta}$
\begin{equation}
{\lambda}_{\eta}=lim_{{t}\rightarrow{\infty}}[sup||\frac{1}{t}log(e^{t{\Gamma}_{\eta}})||]
\label{48}
\end{equation}
where
\begin{equation}
sup\{{\lambda}_{\eta}(x,v):x\in{\cal{M}},v\in{T\cal{M}}\}\le{0}
\label{49}
\end{equation}
One also notes that, the following cosmological constraint is given by
\begin{equation}
R+\frac{1}{2}{\theta}\ge{0}
\label{50}
\end{equation}
This constraint shows that if the Ricci scalar $R>0$ as is expected by decaying, the expansion ${\theta}$ has to be positive as well and the magnetic field decay as the universe expands is consistent with anti-fast dynamo Vishik theorem \cite{3}. It is interesting to observe that these bounds on the Ricci tensor are also important in mixing flows, as has been demonstrated by Thiffeault and Boozer \cite{12}.
\section{Conclusions} Stretching magnetic field lines by plasma flows has led to dynamo action, as shown recently by M Nunez \cite{18} in the case of the nonlinear hydromagnetic dynamos. Here by making use of the spectrum of the fast dynamo operators it is shown that de Sitter-Lyapunov metric in Ricci flows play the same role in two space section of Einstein-de Sitter-Ricci flows cosmology. As pointed out by Woolgar \cite{16} Einstein gravity called general relativity does not show any departure from Ricci flow gravity, however, the same does not happen in Ricci flow cosmology. The example investigated here shows that, though small differences appears in the fast dynamo operator and eigenvalue spectra, these differences actually represents no contradict with the cosmological experiments. Besides the fast dynamo action for de Sitter or closed (2+1)-spacetime Ricci flows, where the cosmological constant ${\Lambda}>0$, which is a new result, one is able to reproduce the BT magnetic field decay in $(2+1)$ real spacetime of GR cosmology. One of the main results of the paper, is that dynamo action is transported along Ricci flow, and that the eigenvalues of fast dynamo operator are also Ricci scalar dependent. For the interested mathematician, a detailed account of GR cosmological dynamos is contained in Widrow \cite{19}. A more detailed account of Einstein manifolds can be seen in Besse book \cite{20}. Last but certainly not least, Thiffeault \cite{21} called my attention to the fact that the type of maps considered here are not actually Anosov, because the Lyapunov exponents in this case are constant everywhere in the manifold, which shall impose severe constraints on the cosmological models in Ricci flows, unless more general Anosov maps can be considered as the pseudo-Anosov map considered by Gilbert \cite{22}, in the case of stretch-fold-shear dynamos. Recent investigation of Vishik anti-fast dynamo theorem \cite{4}, indicates that the magnetic flux tube dynamos in two dimensions possess negative Riemannian curvature. This could be also considered a non-cosmological application of the kinematic dynamo spectra addressed here.\section{Acknowledgements}
Several discussions with Yu Latushkin, D Sokoloff and J L Thiffeault are highly appreciated. I also thank financial  supports from UERJ and CNPq.
 

\begin{thebibliography}{22}
\bibitem{7} G Perelmann, The entropy formula for the Ricci flow and its geometric applications , Los Alamos arxives math-DG /0211159.
\end{thebibliography}
  \end{document}